# Direct determination of Spin-Orbit torque by using dc current-voltage characteristics

R. Guerrero[1*], A. Anadon[1], A. Gudin[1], J. M. Diez[1,2], P. Olleros-Rodriguez[1], M. Muñoz[3], R. Miranda[1,2,4], J. Camarero,[1,2,4] and P. Perna[1*]

[1] IMDEA Nanociencia, c/ Faraday 9, Campus de Cantoblanco, 28049 Madrid, Spain

[2] Dpto. Física de la Material Condensada & Instituto "Nicolás Cabrera", Universidad Autónoma de Madrid, 28049 Madrid, Spain.

[3] IMN-Instituto de Micro y Nanotecnología, (CNM-CSIC), Isaac Newton 8, 28760 Tres Cantos, Madrid, Spain.

[4] IFIMAC, Universidad Autónoma de Madrid, 28049 Madrid, Spain.

[*]*Corresponding Author Email: ruben.guerrero@imdea.org ; paolo.perna@imdea.org*

**Abstract**

**Spin polarized currents are employed to efficiently manipulate the magnetization of ferromagnetic ultrathin films by exerting a torque on it. If the spin currents are generated by means of the spin-orbit interaction between a ferromagnetic and a non-magnetic layer, the effect is known as spin-orbit torque (SOT), and is quantified by measuring the effective fields produced by a charge current injected into the device. In this work, we present a new experimental technique to quantify directly the SOT based on the measurement of non-linearities of the dc current-voltage (IV) characteristics in Hall bar devices employing a simple instrumentation. Through the analysis of the IV curves, the technique provides directly the linearity of the effective fields with current, the detection of the current range in which the thermal effects can be relevant, the appearance of misalignments artefacts when the symmetry relations of SOT are not fulfilled, and the conditions for the validity of the single domain approximations, which are not considered in switching current and second harmonic generation state-of-the-art experiments. We have studied the SOT induced antidamping and field-like torques in Ta/Co/Pt asymmetric stacks with perpendicular magnetic anisotropy.**

The manipulation of the magnetization in ferromagnetic materials through spin-polarized currents is a key technological aspect because it enables applications in fast, high density and low energy consumption storage/processing media. In spin valve structures, in which two ferromagnets (FMs), i.e., free and pinned FMs, are separated by a non-magnetic (NM) material, the magnetization of the free FM is manipulated by exploiting the torque exerted by a spin current generated by the pinned FM and flowing perpendicular to the interfaces. This effect takes place via a spin transfer mechanism (spin transfer torque, STT) [1]. Recently, a different strategy that exploits the spin-orbit coupling (SOC) produced in FM / heavy-metal (HM) interfaces, named as spin-orbit torque (SOT), makes use of an in-plane charge current (parallel to the interface) to efficiently control the magnetization of the FM. This torque is strictly dependent on the nature of the material composing the interface [2], [3]. In most of the technologies proposed, SOT replaces the use of a ferromagnetic material as spin polarizers simplifying notably the architecture of the devices as the induced torque is driven by a charge current flowing in the plane of the NM layer rendering the device more resilient to heating and/or electromigration. SOT has been hence exploited in the writing lines of magnetic random access memories [4], in metallic based nanoscillators [5]–[7] or in magnetic insulators (as yttrium iron garnets) based nanoscillators [8], in which the magnetization dynamics of the FM can be transduced in resistance oscillations either by anisotropic or spin Hall magnetoresistance [9].

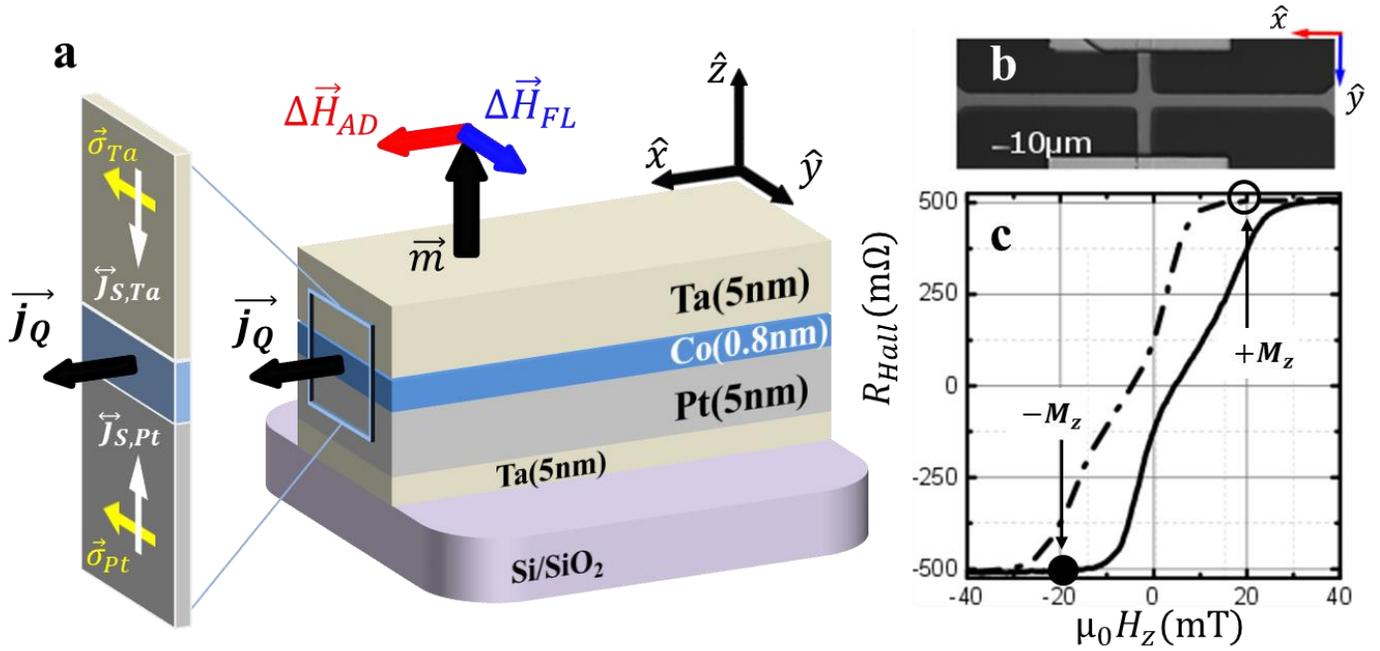

**Figure 1. Schematics of the vectors involved in the SOT phenomena. a)** Sketch of the sample structure, Ta(5nm)/Co(0.8nm)/Pt(5nm)/Ta(5nm)//SiO2(substrate), and of the vectors involved in the SOT phenomena. The current density $\vec{j}_Q$ (black arrow), which is injected along $\vec{x}$, produces a spin current density $\vec{j}_S$ along $\vec{z}$ (white arrows) whose spins are $\vec{\sigma}$ polarized (yellow arrows), that is parallel to $-\vec{y}$. Given the different spin Hall angle in Ta and Pt the spin current generated in the each layer impinges the Co layer with the same polarization. The spin current is responsible of the effective fields, antidamping $\Delta\vec{H}_{AD} = \chi_{AD} j_Q (\vec{m} \times \vec{\sigma})$ (red arrow) and field-like $\Delta\vec{H}_{FL} = \chi_{FL} j_Q \vec{\sigma}$ (blue arrow) torques, acting on the magnetization $\vec{m}$ (black arrow) of the system. **b)** Optical microscope image of the Hall bar device employed in this study, having 10 μm wide channel. Note that the in-plane $x$ ($y$) direction is parallel (perpendicular) to the current density. **c)** Anomalous Hall Effect measurement of the Ta(5nm)/Co(0.8nm)/Pt(5nm) asymmetric stack. Empty (filled) black circle corresponds to a positive (negative) magnetization state labeled as $+M_z$ ($-M_z$).

SOT-effects in perpendicular magnetic anisotropy (PMA) layers in contact with non-magnetic heavy metal (HM) layers hold promise for improved energy-efficiency devices. In fact, SOT can be used to control the displacement of domain walls (DWs) [10]–[14] or switch the magnetization of the FM by a charge current. Besides, it can be exploited to nucleate and move chiral DWs and magnetic skyrmions [15]–[18] in systems with large Dzyaloshinskii-Moriya interaction [19]–[21], enabling the realization of many promising technological applications such as skyrmion based racetrack memories.

The schematics of the SOT phenomena in asymmetric trilayers HM1/FM/HM2, as the one investigated here, is shown in **Figure 1a**. By applying a planar charge current ($\vec{j}_Q$) along $\vec{x}$ through the HM layers (gray and brown layers), the magnetization of the FM (blue layer) experiences a torque due to the spin current ($\vec{j}_S$) flowing along $\vec{z}$ and comprising of spins $\vec{\sigma}$ polarized. $\vec{j}_S$ raises from the interfacial and/or the bulk SOC by means of spin Hall effect characterized by the spin Hall angle of the HMs. The opposite spin Hall angles of Pt and Ta, used as bottom and top interfaces, promote the same $\vec{\sigma}$, pointing along -$\hat{y}$. SOT is thus identified by two effective fields, namely $\mu_0 \Delta \vec{H}_{AD}$ and $\mu_0 \Delta \vec{H}_{FL}$, which are due to the antidamping and field-like torques $\chi_{AD}$ and $\chi_{FL}$, respectively. These fields vary linearly with the charge current density, i.e. $\mu_0 \Delta H_{AD/FL} = \chi_{AD/FL}\, j_Q$, and are modulated by the $\chi_{AD/FL}$ parameters, which are the main figures of merit of SOT.

The most used techniques employed for the estimation of the effective fields $\mu_0 \Delta H_{AD/FL}$, from which the torques are deduced, are based on the measurements of the switching current density [22] and of the second harmonic (SH) Hall voltages [23]. However, both techniques present undesirable features. In the former case, the torque is estimated by measuring the critical currents that promote the perpendicular magnetization switching as function of in-plane fields under the assumption of coherent rotation magnetization mechanisms (macro spin modelling) even though this is far from the irreversible nature of the magnetization transitions. The SH experiments rely on how the effective fields (generated by alternate current (ac)) compete with an external applied field and hence modify the second harmonic signals (measured via lock-in techniques) [23]-[27]. The SH technique is based on the assumption of a linear dependence of the effective fields with the charge current injected into the device, which is used to extract the torque parameters $\chi_{AD/FL}$.

In this work, we report on a new experimental technique for the determination of $\chi_{AD}$ and $\chi_{FL}$ by quantifying the non-linearities in the anomalous Hall resistance via the measurement of dc current-voltage (*IV*) characteristics. The method demands currents of the order of $10^{10} Am^{-2}$, and requires simple current source-voltmeter instead of lock-in amplifiers employed to detect harmonic signals. Through the analysis of the *IV* curves, our technique provides directly the linearity of the effective fields with current, the detection of the current range in which the thermal effects can be relevant, the symmetry relations of SOT that reveal the appearance of misalignments artefacts, and the conditions for the validity of the single domain approximations, which have been never discussed previously. Both the experimental procedure and the protocol analysis for quantifying the SOT are provided. Finally, we demonstrate that the data obtained by using our technique and the second harmonic are in agreement.

The experiments were performed in asymmetric trilayer stacks grown onto Silicon (Si) substrates capped by 300 nm thick $SiO_2$. The complete structure is Ta(5nm)/Co(0.8nm)/Pt(5nm)/Ta(5nm)//$SiO_2$(300nm)/Si(substrate). The layers were deposited by DC sputtering in $8 \times 10^{-3}$ mbar Ar partial pressure (base pressure of the sputtering chamber was $10^{-8}$ mbar) with a deposition rate of 0.3 Å/s (monitored by a quartz microbalance) [28], [29]. The sample was then exposed to air to oxidize the Ta top-most layer. To define the Hall bar devices with 10μm Hall bars we used optical lithography (photoresist AZ1512) and Ar milling etching. Ta(10nm)/Cu(100nm)/Pt(10nm) contacts were grown on top of the structure by DC sputtering in a second lithography step (Photoresist AZ1512). In **Figure 1b**, an optical microscope photograph shows the obtained device. By injected electric current $j_Q$ along $\hat{x}$ and by measuring the Hall voltage ($V_{Hall}$) along $\hat{y}$ as function of the magnetic field applied along the $\hat{z}$ direction, we obtain the anomalous Hall resistance ($R_{Hall}$) hysteresis curve of the device (**Figure 1c**), which displays the typical behavior of a PMA system.

To quantify $\chi_{AD}$ and $\chi_{FL}$ of the stack we measured the *IV* curves of the Hall resistance as function of a constant dc in-plane magnetic field, applied either along $\hat{x}$ or $\hat{y}$ direction starting from both positive and negative saturation states. In the former case, the procedure adopted in the experiments is the following. *First*, the sample was saturated by applying +40mT along +$\hat{z}$ axis, and then $H_z$ was kept to +20mT that corresponds to a magnetization state close to the positive saturation (labeled as +$M_z$ in Figure 1c) for the different in-plane fields used. In the other case, the sample was first saturated negative, using $H_z = -40$mT, and then kept at the negative saturation state by using $H_z = -20$mT (corresponding to a magnetization state labeled as $-M_z$ in Figure 1c). This method ensures that the sample magnetization rotates reversibly around its positive (i.e., +$M_z$) or negative (i.e., $-M_z$) saturation state during the in-plane magnetic sweeps (schematically sketched in the inset of Figure 2a), avoiding irreversible domain nucleation-propagation processes. *Second*, we measured *I-V* curves at different in-plane field values of $\mu_0 H_{x,y}$. By measuring from the two opposite saturation conditions, the current-driven SOT phenomena in the asymmetric PMA trilayers are double checked. Moreover, a simple analysis of the symmetry of the experimental data provides information on artefacts originated from instrumental misalignments. These are discussed in the following. From each *I-V* ($\mu_0 H_{x,y}$) curve acquired starting from +$M_z$ and $-M_z$, we extracted a linear and a quadratic term (see details in Sections 1 and 2 of the Supplemental Information). In brief, the Hall resistance ($R_{Hall}$) is obtained by performing a linear fit of the *I-V* data, whereas the even part of the Hall voltage ($V_{Hall}$), defined as $V_{even}(I) = \frac{V_{Hall}(I) + V_{Hall}(-I)}{2}$, provides the SOT-driven torques. **Figure 2a** shows the measured $R_{Hall}$ as function of $\mu_0 H_x$ in the low current range (i.e., below < 500μA that corresponds to $j_Q$=8.3x10⁹ Am⁻²). Positive (negative) values of the resistance refer to $+M_z$ ($-M_z$) magnetization states. **Figure 2b,c** displays the dependences of $V_{even}$ with the current density at different values of $\mu_0 H_x$ for $+M_z$ and $-M_z$, respectively. From a simple inspection of panels b and c, we observe a quadratic dependence of $V_{even}$ on the applied current for both magnetization states. Similar trends, but with smaller variations, are found when the field is applied perpendicular to current flow, i.e., for $\mu_0 H_y$ (see Figure S2). Therefore, the experimental data directly prove the quadratic dependence of $V_{even}$ with the injected current, i.e. $V_{even} \propto \mu_0 H_{x,y} \, j_Q^2$.

Remarkably, the experimental data display well-defined symmetry when current, fields and device are well aligned. For instance, the Hall resistance has specular symmetries with respect to the field direction and the magnetization state, i.e., $R_{Hall}(+M_z) = -R_{Hall}(-M_z)$ and $R_{Hall}(+H_{x,y}) = R_{Hall}(-H_{x,y})$, whereas $V_{even}$ at fixed injected current varies with in-plane fields as $V_{even}(j_Q, +H_{x,y}) = -V_{even}(j_Q, -H_{x,y})$. The absence (even partial) of these symmetries indicates the existence of instrumental misalignments. Although the partial loss of the symmetry does not affect the quantification of the torques (see Section S4 of the Supplementary Information), our method based on dc current-voltage characteristics provides direct fingerprints to detect experimental misalignments.

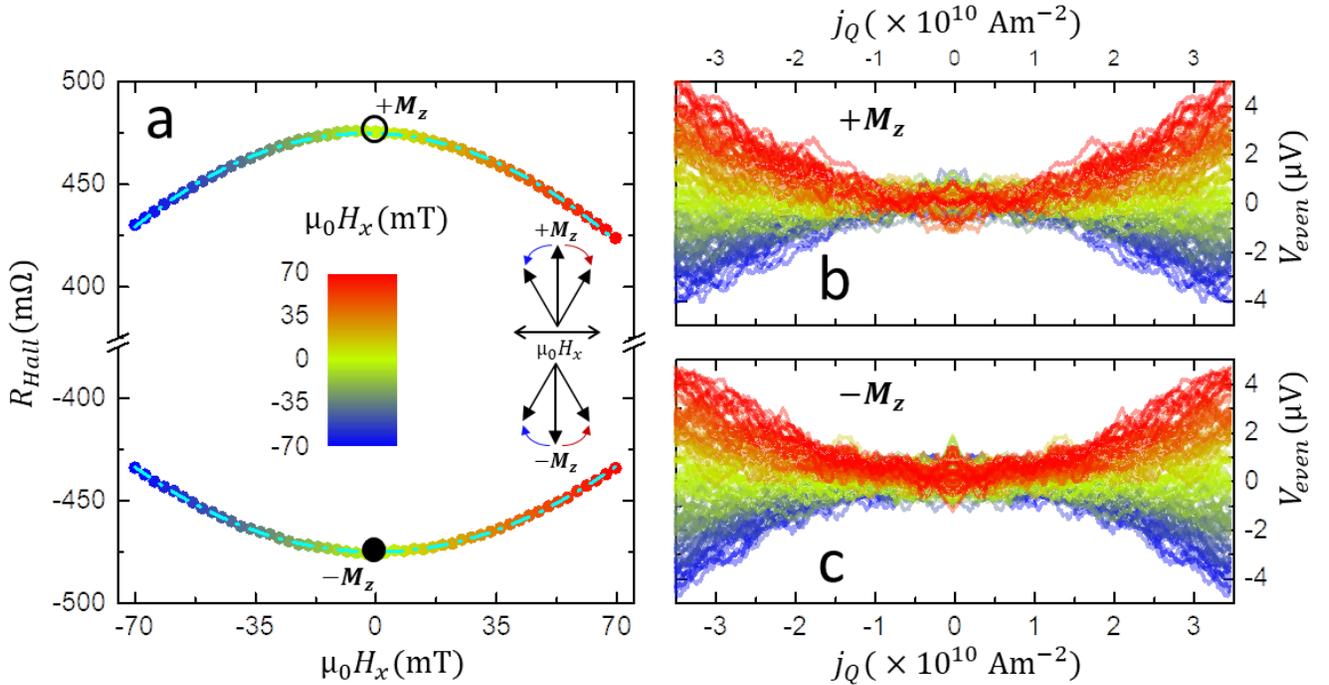

**Figure 2. Quadratic dependence of $V_{even}$ with the current density, $j_Q$. a)** Hall resistance extracted from *I-V* curves obtained at different in-plane magnetic fields oriented along the $\hat{x}$ direction, i.e., parallel to the current flow. The top (bottom) curve corresponds to the magnetization state, labeled as $+M_z$ ($-M_z$), close to the positive (negative) saturation state (i.e., corresponding to the empty and filled circles in Figure 1c). During the in-plane magnetic sweeps, the sample magnetization rotates reversibly around $+M_z$ or $-M_z$ (as sketched in the inset). **b,c)** Even part of the *I-V* curves obtained at different $\mu_0 H_x$ at $+M_z$ (panel b) and $-M_z$ (panel c) magnetization states. The color code indicates the applied magnetic field values. Similar behavior is obtained as function of $\mu_0 H_y$ (see Figure S2 in Supp. Info.). From each curve the quadratic coefficient of $V_{even}(j_Q, \mu_0 H_{x,y})$ , i.e., $C_{even}(\mu_0 H_{x,y})$, that yields to the SOT effective fields, is extracted by performing a parabolic fit.

To determine the $\chi_{AD,FL}$, we study the dependence of $V_{Hall}$ with the in-plane magnetic field. From [23], we can express z-component of the magnetization:

$$m_z \sim \pm \left[1 - \frac{1}{2}\left[\frac{(\mu_0 H_{x,y} + \mu_0 \Delta H_{AD,FL}(j_Q))}{\frac{2K_u}{M_s} - \mu_0 M_s + \mu_0 H_z}\right]^2\right] \quad (1)$$

Where, $M_s$ is the saturation magnetization and $K_u$ is the anisotropy energy. This expression is valid in whole field range investigated, as the solid lines of **Figure 2a** shows, i.e., for relatively small magnetization rotation values close to the starting magnetization state, as it is a second order series expansion. Since $V_{Hall} = S j_Q R_{AHE,max} m_z$, and for a given in-plane field condition, the dependence of the Hall voltage with the injected current can be expressed simply with linear and quadratic coefficients (see Sections 2 and 3 of the Supplemental Information), which can be directly related with the odd and even parts of the experimental data:

$$V_{Hall}(\mu_0 H_{x,y}, j_Q) \sim C_{odd}(\mu_0 H_{x,y}) j_Q + C_{even}(\mu_0 H_{x,y}) j_Q^2 = V_{odd}(j_Q) + V_{even}(j_Q) \quad (2)$$

in which the coefficients are defined as

$$C_{even}(\mu_0 H_{x,y}) = \mp S R_{AHE,max} \left[\frac{2\mu_0 H_{x,y} \chi_{AD,FL}}{\left[\frac{2K_u}{M_s} - \mu_0 M_s + \mu_0 H_z\right]^2}\right]$$

$$C_{odd}(\mu_0 H_{x,y}) = \pm S R_{AHE,max} \left[1 - \left\{\frac{[\mu_0 H_{x,y}]^2}{\left[\frac{2K_u}{M_s} - \mu_0 M_s + \mu_0 H_z\right]^2}\right\}\right]$$

To note that both $\chi_{FL}$ and $\chi_{AD}$ are included in the quadratic term of $j_Q$. $C_{even}$ can be hence extracted by a parabolic fit of the measured $V_{even}(j_Q)$. It is worth noting that the parabolic dependence of $V_{even}$ with $j_Q$ implies the linearity of $\mu_0 \Delta H_{AD/FL}$ with current density.

The top graphs of **Figure 3** compare the dependence of $C_{even}$ with the in-plane magnetic field parallel (a) and perpendicular (b) to the current flow direction. In both cases $C_{even}$ evolves linearly with the in-plane magnetic field, so that $\chi_{AD,FL}$ can be obtained from the slope of the curve normalized by the second derivative of $C_{odd}$ with respect to $H_{x,y}$, i.e.:

$$\chi_{AD,FL} = \frac{|\partial C_{even}/\partial H_{x,y}|}{|\partial^2 C_{odd}/\partial H_{x,y}^2|} \quad (3)$$

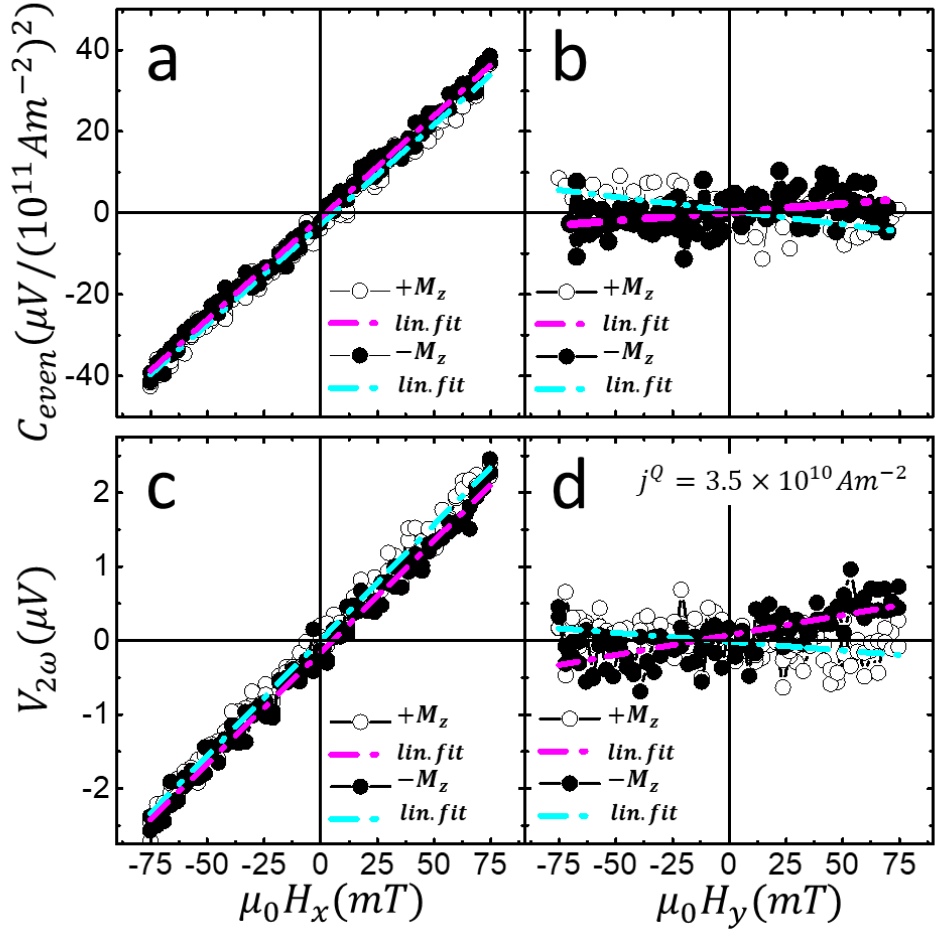

**Figure 3. Experimental determination of SOT** by dc current-voltage characteristics (panels a and b) and by second harmonic (SH) voltage analysis (panels c and d) derived from the applied in-plane field dependences of their characteristic parameters $C_{even}$ and $V_{2\omega}$, respectively. The $C_{even}$ values are derived from the parabolic dependence of the *I-V* curves acquired at different in-plane fields, as the ones shown in Figure 2b. The $V_{2\omega}$ values come from the FFT analysis of the corresponding *I-V* curves from other experimental dataset for a given current density. The symbols correspond to the experimental parameters derived from the dc and SH analysis. Empty and filled symbols have been used to distinguish between $+M_z$ and $-M_z$ magnetization states. The solid lines are linear fits to yield to the effective fields due to SOT.

yielding to $\chi_{AD}$=(4.5±0.5) mT/($10^{11}$Am$^{-2}$) and $\chi_{FL}$=(0.6±0.1) mT/($10^{11}$Am$^{-2}$).

Some aspects deserve for attention. *First*, the out-of-plane field ($H_z = \pm 20$mT) used to ensure that the magnetization rotates reversibly around its positive or negative ($\pm M_z$) saturation state during the in-plane magnetic sweeps, which avoid irreversible domain nucleation-propagation processes, is included in the expression for $C_{even}$ and $C_{odd}$ (Eq. 2), generalizing the application of the method to systems with different PMA. *Second,* although not drastically affecting the estimation of the $\chi_{AD/FL}$, slight geometrical misalignments (discussed in Section 4 of the Supplementary Information) between the sample's surface plane, the applied in-plane field and the injected current direction, which may be also present in SH experiments, introduce an offset in $C_{even}$ (i.e., deviation from $C_{even} \neq 0$ at $\mu_0 H_{in-plane} = 0$) and alter the symmetry relations $C_{even}(H_x, +M_R) = C_{even}(H_x, -M_R)$ and $C_{even}(H_y, +M_R) = -C_{even}(H_y, -M_R)$. *Third*, while from the *I-V* experimental curves obtained in our method it is possible to retrieve the second harmonic signal generated by SOT (by using the Fourier transformation), from the SH measurements it is not be possible to obtain the *I-V* curves. In fact, in SH experiments, an ac sinusoidal current is applied to the device and the voltage output is measured using a lock-in amplifier, which enables the analysis of first ($V_{1\omega}$) and second ($V_{2\omega}$) order harmonic Hall voltages. The first term, $V_{1\omega}$, is related to $m_z$, whereas $V_{2\omega}$ contains the information of the torques, similarly to the parameters $R_{Hall}$ and $C_{even}$ in our dc measurements. Figure 3c and Figure 3d show the field evolutions of $V_{2\omega}$ for a current density of 3.5x$10^{10}$Am$^{-2}$ with $\mu_0 H_x$ and $\mu_0 H_y$, respectively. The antidamping and field-like torques have been then extracted by using [23]:

$$\mu_0 \Delta H_{AD,FL}(j_Q) = j_Q \chi_{AD,FL} = 2 \frac{|\partial V_{2\omega}/\partial H_{x,y}|}{|\partial^2 V_{2\omega}/\partial H_{x,y}^2|}, \quad (4)$$

yielding to $\chi_{AD}^{2\omega}$=(6.0±0.6) mT/($10^{11}$Am$^{-2}$) and $\chi_{FL}^{2\omega}$=(0.7±0.3) mT/($10^{11}$Am$^{-2}$), which are very close to the values obtained by our method.

Finally, it is worth mentioning that the measured signals have to be corrected by the planar Hall effect (PHE) [27], while thermoelectric effects, such as Nernst effect, generated by the high current densities, can be removed only at high magnetic fields [25]. The procedure commonly adopted to correct the antidamping and field-like torques due to PHE [30], is to express them in terms of the ratio $\xi = R_{PHE,max}/R_{AHE,max}$ as $\chi_{AD,FL,c} = \frac{\chi_{AD,FL} + 2\xi \chi_{FL,AD}}{1 - 4\xi^2}$. From an independent measurement of the PHE resistance ($R_{PHE}$), obtained by measuring $R_{Hall}$ at different angles from -30 to 30 degrees (around the $\hat{x}$ direction), we get $\xi = 0.35$ (see Section 4 of the Supplemental information). The PHE corrected torques are hence: $\chi_{AD}$=(9±1) mT/($10^{11}$Am$^{-2}$) and $\chi_{FL}$=(6.4±0.7) mT/($10^{11}$Am$^{-2}$). Same corrections can be applied to the values of the effective fields obtained using the second harmonic technique, bringing to similar values: $\chi_{AD}^{2\omega}$=(12±1) mT/($10^{11}$Am$^{-}$

$^2$) and $\chi_{FL}^{2\omega}$=(9±1) mT/(10$^{11}$Am$^{-2}$). These results are in line with those obtained by Woo *et al.* [31] for similar structures (small differences may be due to slightly different composition or layers thickness of the stacks).

In conclusion, we have developed a new technique to determine the SOT induced antidamping and field-like effective fields in PMA stacks of interest for fast, efficient and energy saving spin-orbitronic technological applications. The technique exploits the non-linearities induced by the spin torque in the resistance output of the system, and only requires dc measurements of the current-Hall voltage characteristics as function of an externally applied magnetic field parallel/perpendicular to the current density vector. The even part of such curves allows for the direct determination of the antidamping and field-like torques, from which the corresponding effective fields and torques are derived. We have employed this technique to quantify the SOT in Ta/Co/Pt asymmetric trilayers. The results obtained have been compared with second harmonic analysis yielding equivalent results. Our technique provides directly the linearity of the effective fields with current, the detection of the current range in which the thermal effects can be relevant, the appearance of misalignments artefacts, and the conditions for the validity of the single domain approximations. The developed technique determining directly the SOT and its dependence with current in magnetic structures avoiding the use of more complex ac techniques.

**ACKNOWLEGDEMENTS**

This research was supported by the Regional Government of Madrid through Project P2018/NMT-4321 (NANOMAGCOST-CM), by the Spanish Ministry of Economy and Competitiveness (MINECO) through Projects RTI2018-097895-B-C42 (FUN-SOC), FIS2016-78591-C3-1-R (SKYTRON), MAT2017-87072-C4-4-P (HEDIMAG), and by the PCI2019-111867-2 (FLAG ERA 3 grant SOgraphMEM). JMD and AG acknowledges support through MINECO and CM through grants BES-2017-080617 and PEJD-2017-PREIND-4690, respectively. IMDEA Nanoscience is supported by the 'Severo Ochoa' Programme for Centres of Excellence in R&D, MINECO [grant number SEV-2016-0686]. We acknowledge M. R. Osorio and D. Granados at IMDEA-nanoscience nanofabrication center for their help in the lithography process.

**Corresponding authors**

ruben.guerrero@imdea.org ; paolo.perna@imdea.org